\documentclass{article}
 \usepackage{graphicx,cite}

\def\ss{\relax{\supset\kern-13pt +}}
 \title { Lie Point Symmetries and Commuting Flows for Equations on Lattices }
 
\author{ D. Levi\\Dipartimento di Fisica "E. Amaldi", \\ Universit\'a degli
 Studi Roma Tre and Sezione INFN, Roma Tre,\\Via della Vasca Navale 84,
 00146 Roma, Italy
 \and P. Winternitz\\
Centre de recherches math\'ematiques and \\D\'epartement de math\'ematiques et 
de statistique,\\
Universit\'e de Montr\'eal\\ 
C.P. 6128, succ. Centre-ville,\\
Montr\'eal (Qu\'ebec), H3C 3J7, Canada}

 \date{\today}
 
 \def\pa {\partial}
 
 \def\ra {\rightarrow}
 
 \def\eq {\equiv}

 \def\be   {\begin{equation}}   \def\ee   {\end{equation}}
 \def\ba   {\begin{array}}      \def\ea   {\end{array}}
 \def\bea  {\begin{eqnarray}}   \def\eea  {\end{eqnarray}}
 \def\bean {\begin{eqnarray*}}  \def\eean {\end{eqnarray*}}

\begin{document}
 \maketitle
 
 \begin{abstract}
 Different symmetry formalisms for difference equations on lattices are reviewed 
and applied to perform symmetry reduction for both linear and nonlinear partial 
difference equations. Both Lie point symmetries and generalized symmetries are 
considered and applied to the discrete heat equation and to the integrable 
discrete time Toda lattice.

 \centerline {\bfseries  R\'esum\'e}
 
 Deux formalismes diff\'erents pour \'etudier les sym\'etries des \'equations 
aux  diff\'erences finies sur un r\'eseau sont d\'ecrits et utilis\'es pour 
faire la r\'eduction par sym\'etrie des \'equations aux diff\'erences finies.
 Les sym\'etries ponctuelles et g\'en\'eralis\'ees sont consid\'er\'ees et 
appliqu\'ees \`a l'\'equation de la chaleur lin\'eaire discr\`ete et \`a un 
treillis de 
Toda int\'egrable en temps discret. 
  \end{abstract}

 \section{Introduction}

The purpose of this article is to compare two different approaches to the 
study of symmetries of difference equations 
\cite{1,2,3,4,5,6,7,8,9,10,11,12,13,14,15,16,17,18,19,20,21,22,23,20c,20a,13d,13
e,13c,20b,25}. Both 
approaches are algebraic, in that they both use an infinitesimal formalism in 
which certain vector fields realize a Lie algebra: the symmetry algebra. In one 
approach the vector fields act on the dependent and independent variables and 
in general on the difference equations and the lattices. In the other approach, 
use is made of evolutionary vector fields, acting only on the dependent 
variables. In both approaches, further choices must be made. In particular, 
the coefficients of the vector fields can depend on the dependent and 
independent variables at one point of the lattice, on a finite number of points, 
or on an infinite one. They may also depend on derivatives of the dependent 
functions, up to some chosen order, or on finite dfferences.

For point symmetries of {\sl differential} equations the two approaches are 
equivalent and the formalisms of ordinary vector fields and evolutionary ones 
are related in a simple manner. In particular, the evolutionary formalism 
provides flows commuting with the original equation. If these flows involve 
first derivatives only, and they figure linearly, they correspond to point 
symmetries. The two formalisms provide the same symmetry variables and hence the 
same symmetry reductions.

For {\sl difference} equations the situation is somewhat different. Also in 
this case, for point symmetries it is very natural to consider vector fields 
involving differentiation with respect to both dependent and independent 
variables. Moreover, it is natural to let the transformations act on the 
difference equation, and on the lattice itself.

A purely difference equation does not contain derivatives (by definition). 
Hence the usual evolutionary formalism, with the characteristic $Q$ of the 
symmetry depending on $x$, $u$ and derivatives of $u$, is not a natural tool to 
use. The natural evolutionary formalism for discrete equations is one in which 
$Q$ depends on the values of the variables $x$ and $u$ at different points of 
the lattice. Such evolutionary symmetries are in general not point symmetries 
and are not related to point symmetries in any simple manner.

One of the most important applications of symmetries of differential equations 
is to perform symmetry reduction. For partial differential equations this means 
a reduction of the number of independent variables.

In this article we shall perform symmetry reduction for various partial 
difference equations and show how different types of symmetries lead to 
different results.

The general theory is discussed in Section 2, first for differential 
equations, then for difference ones. Also in Section 2 we introduce the concept 
of "discrete evolutionary vector fields" and their prolongations. Section 3 is 
devoted to symmetry reduction for a linear difference equation, namely the 
discrete heat equation. Translations, as point symmetries, provide a reduced one 
variable equation that is easily solved. Dilations provide a reduction to a
dilation-delay equation. Still in Section 3 we use discrete evolutionary vector 
fields to obtain solutions of the discrete heat equation, invariant under 
translations and under dilations. Similarly, in Section 4, we use first point 
symmetries, then discrete evolutionary vector fields to obtain reductions of an 
integrable nonlinear partial  difference equation: the discrete time Toda 
lattice.

  \section{Continuous Symmetries of Difference Schemes}

  \subsection{Equivalence of two symmetry formalisms for differential equations}

 Let us first recapitulate a well-known result for differential equations, 
namely that Lie point symmetries (and also generalized symmetries) can be 
realized in two different ways. The first is by vector fields acting on both 
independent and dependent variables, the second by evolutionary vector fields, 
acting on the dependent variable only. This is true for arbitrary systems of 
differential equations (ordinary or partial, of any order) \cite{24}. For point 
symmetries the two realizations are entirely equivalent. For generalized 
symmetries ordinary vector fields can always be replaced by evolutionary ones in 
a straightforward manner. Let us illustrate this equivalence on the example of
 a first order ordinary differential equation
\be \label{2.1}
E \eq u_x - F(u,x) = 0.
\ee
Let us assume that
\be \label{2.2}
\hat X = \xi(x,u) \pa_x + \phi(x,u) \pa_u
\ee
generates a point symmetry, i.e satisfies
\be \label {2.3}
pr\hat{X} E|_{E=0} = 0.
\ee
The first prolongation of $\hat{X}$ is
\be \label{2.4}
pr \hat X= \hat X + \phi^x\pa_{u_x},\qquad \phi^x =D_x\phi - (D_x\xi) u_x ,
\ee
where $D_x$ is a total derivative.

The prolonged evolutionary vector field, on the other hand, is
\be \label{2.5}
pr \hat{X_e} = Q \pa_u + Q^x \pa_{u_x}, \quad   Q^x = D_x Q,
\ee
where $Q$ is the characteristic of the vector field.

The total derivative $D_x$ is itself a generalized symmetry of eq. (\ref{2.1})
(and of any equation). Indeed, if we have $E=0$, then 
$D_x E = 0$ follows and so does $(\xi D_x) E=0$. Hence, if $\hat X$ satisfies
eq. (\ref{2.3}) we also have
\be \label {2.6}
(pr \hat X - \xi D_x) E|_{E=0} = 0.
\ee

In view of eq. (\ref{2.4}) we have
\bea \label {2.7}
pr \hat X- \xi D_x =& \xi \pa_{x} + \phi \pa_u + [ D_x\phi - (D_x \xi) u_x] 
\pa_{ u_x} -
\\ \nonumber
&-\xi[\pa_x + u_x\pa_u +u_{xx}\pa_{u_x}]=
\\ \nonumber
=&[\phi - \xi u_x] \pa_u + [D_x (\phi - \xi u_x)] \pa_{u_x}.
\eea
With
\be \label{2.8}
Q\equiv \phi-\xi u_x, 
\ee
we obtain eq (\ref{2.5}) and we have proved that for first order ODEs condition 
(\ref {2.3}) is equivalent to the condition
\be \label {2.9}
\hat X_e E|_{E=0} = 0,
\ee
i.e. that the two formalisms are equivalent. The same is true for higher order 
ODEs and PDEs of any order \cite{24}.

Given a vector field (\ref{2.2}) we can calculate the corresponding 
characteristic $Q$ and the evolutionary field $\hat X_e$ of eq. (\ref{2.5}). 
Hence
flows that commute with the flow given by the considered equation have the form
\be \label{2.10}
u_{\lambda} = Q(x,u,u_x,u_{xx},...),
\ee
where $\lambda$ is a group parameter.

\subsection{Point symmetries of difference equations}

	In two recent articles \cite {10,12} we presented a method for 
determining Lie point symmetries of difference systems. In this terminology a 
"difference system" is a system of relations between a set of points in a 
$(p+q)$ dimensional space where the $p$ coordinates $x_1,...,x_p$ represent 
independent variables and the $q$ coordinates $u_1,...,u_q$ represent the 
dependent ones. Let us restrict here to the case of one difference equation for 
one function $u$ of one variable $x$. Let us consider a simple case when when 
three points
$ P, P^+,P^- $ are involved. 

These three points have coordinates $(x,u)$, $(x^-,u^-)$, and $(x^+,u^+)$ , 
respectively. The difference system consists of two relations
\be \label{2.11}
E_a (x,x^-,x^+,u,u^-,u^+) = 0, \quad a=1,2.
\ee

This system describes both a difference equation and a lattice. If a continuous 
limit $x^+\rightarrow x$ , $x^-\rightarrow x$ exists, then one of these 
equations goes into a first order differential equation, the other into an 
identity (like $0=0$).

The algorithm for finding the Lie point symmetries of such a system is quite 
simple \cite{10,12}. We write a vector field as in the continuous case and 
prolong it to all points involved in the difference scheme. In the case of 
system (\ref{2.11}) we have
\bea \label {2.12}
pr\hat{X} &=&\xi(x,u) \pa_x +\phi(x,u)\pa_u +\xi(x^+,u^+)\pa_{x^+}+
\\ \nonumber
&+&\phi(x^+,u^+)\pa_{u^+} +\xi(x^-,u^-)\pa_{x^-} +\phi(x^-,u^-)\pa_{u^-}.
\eea

The algorithm for determining the functions $\xi(x,u)$ and $\phi(x,u)$ is 
\be \label {2.13}
pr\hat{X} E_a|_{E_i=0}=0, \quad  a=1,2,\quad  i=1,2.
\ee
The algorithm provides functional equations for $\xi$ and $\phi$. Solution 
methods were discussed elsewhere, as were applications \cite{10,12}.

\subsection{Commuting flows and evolutionary symmetries for\\ difference 
equations}

Let us again consider a three point ordinary difference system like that of eq. 
(\ref{2.11}), but in a form explicitly solved for $u^+$ and $x^+$:
\be \label {2.14}
E_1 = u^+ - f_1(x,x^-,u,u^-)=0, \quad E_2 = x^+ - f_2(x,x^-,u,u^-) =0.
\ee

A Lie point symmetry (\ref{2.12}) can be converted into an evolutionary one in 
the same way as was done in the continuous case. Indeed we can define an
operator ${\hat X}_e$ and its prolongation $pr{\hat X}_e$ by putting
\bea \label {2.15}
pr{\hat X}_e\equiv pr\hat{X} -\xi D_x -\xi^+ D_{x^+} - \xi^- D_{x^-}
\\ \nonumber 
\xi =\xi(x,u),\quad \xi^{\pm} \equiv \xi(x^{\pm}, u^{\pm}),
\eea
where $D_x$, $D_{x^+}$, and $D_{x^-}$ are total derivatives with respect to $x$,
$x^+$ and $x^-$ , respectively. This is equivalent to defining an evolutionary 
symmetry for difference equations in the same way as for differential ones, 
namely
\be \label {2.16}
{\hat X}_e =[ \phi(x,u) - \xi(x,u) u_x] \pa_{u}
\ee
and its prolongation as
\bea \label {2.17}
pr{\hat X}_e = (\phi - \xi u_x) \pa_{u} + (\phi^+ -\xi^+ u^+_{x+})\pa_{u^+}+
\\ \nonumber
+( \phi^- - \xi^- u^-_{x^-}) \pa_{u^-},
\eea
where the superscripts $+$ and $-$ correspond to total shifts for the 
corresponding variables and functions. Using the operator (\ref{2.16}) is 
equivalent to using operator (\ref {2.12}) for point symmetries. The operator
(\ref{2.16}) is not a convenient one for generalizations going beyond point 
symmetries.

Indeed, for difference equations it is more natural (and more fruitful) to 
consider evolutionary vector fields with characteristics $Q$ that depend on the 
independent and dependent variables at different points on the lattice, rather 
than on derivatives, as in eq. (\ref {2.16}) and (\ref{2.17}). In this case we 
have
\bea \label {2.18}
{\hat X}_e = & Q( T^k x, T^k u) \pa_{u}
\\ \nonumber
pr{\hat X}_e = & Q(T^k x,T^k u)\pa_{u} + (TQ)\pa_{u^+} + (T^{-1} Q)\pa_{u^-}              
\\ \nonumber
& m \le k \le n, \quad m,n \in  Z.
\eea

Thus $Q$ depends on $x$ and $u$ at a finite, or possibly infinite number of 
different points. In (\ref{2.18}) $T$ is a total shift operator:
\bea \label {2.19}
Tx= x^+,\quad T^{-1}x = x^-, \quad Tu(x)=u(x^+) \eq u^+,\quad T^{-1}u(x)=u(x^-) 
\eq u^-
\eea

The symmetry algorithm is 
\be \label {2.20}
pr{\hat X}_e E_a |_{E_b=0} = 0,\quad a,b =1,2.
\ee
(together with $E_b=0$ we must also use all shifted equations like $T^k E_b 
=0$).

This is equivalent to requesting that the flow
\be \label {2.21}
u_{\lambda} = Q(T^kx, T^ku), \quad x_ {\lambda} = 0
\ee
should commute with the flow (\ref{2.14}).

To sum up, for difference equations we consider two different algebraic symmetry 
formalisms. The first uses ordinary vector fields of the form (\ref{2.2}), 
coinciding with those used for differential equations. Their prolongation is 
different, see eq. (\ref{2.12}). The algorithm for determining the coefficients 
of the vector fields is given in eq. (\ref{2.13}). These vector fields can be 
integrated to provide genuine point transformations taking solutions of 
difference equations into solutions. The set of Lie point symmetries of a 
difference system is in general much more restricted than that of a differential 
one. This is specially true if we consider a difference equation on a fixed 
lattice, i.e when the lattice equation is just $E_2=x^+ - x = \sigma$ where
$\sigma$ is a fixed and nontransforming constant.

The second formalism for symmetries of difference equations uses the "discrete"
evolutionary vector fields (\ref{2.18}) They correspond to generalized 
symmetries for difference systems. They provide quite general commuting flows.
They can be used to perform symmetry reduction. As in the case of differential 
equations, generalized symmetries are particularly useful for identifying 
systems that are integrable, that is those for which a discrete Lax pair exists. 
For linear difference equations these "discrete" evolutionary vector fields
provide commuting difference operators \cite {4,19}.

We demonstrate below, in Sections 3 and 4, that both types of symmetries of 
difference systems are useful for solving difference equations and that they 
provide different types of results.

\section{The Discrete Heat Equation and its Reductions}
 
\subsection{\bf Evolutionary and point symmetries for the heat equation}

Like the continuous heat equation, the discrete one serves as an excellent 
example of the application of group theoretical techniques. Here we shall use it 
to demonstrate the difference between "discrete" evolutionary and point 
symmetries for linear difference equations.

The discrete heat equation can be written as
\be \label{3.1}
\Delta_t u - \Delta_{xx} u = 0
\ee
The discrete derivatives used in Ref. \cite{4} were defined as
\be \label{3.2}
\Delta_t = \frac{T_t - 1}{\sigma_t}, \qquad \Delta_{xx} = \frac{T_x^2 - 2 T_x 
+1}{\sigma_x^2}
\ee
( a better notation would have been $\Delta_t^+$ and $\Delta_{xx}^{++}$).
Here $T_t$ and $T_x$ are shift operators and $\sigma_t$ and $\sigma_x$ are the 
steps in the $t$ and $x$ directions, respectively. The lattice is fixed, uniform 
and orthogonal.  
The Lie algebra of "discrete" evolutionary symmetries was found to be six 
dimensional (after an infinite dimensional subalgebra corresponding to the 
linear superposition principle was factor out).  It is isomorphic to that of the 
continuous heat equation, and was realized by the "discrete" evolutionary vector 
fields
 \bea \nonumber
 P_0  =  (\Delta_t u)\pa_u, \quad P_1 = (\Delta_x u)\pa_u, \quad W = u \pa_u,
 \\ \nonumber
 B  =  ( 2 t T_t^{-1} \Delta_x u + x T_x^{-1} u + \frac{1}{2} \sigma_x T_x^{-1} 
u)\pa_u,
 \\ \label{3.3}
 D  =  [ 2 t T_t^{-1} \Delta_t u + x T_x^{-1} \Delta_x u + ( 1 - \frac{1}{2} 
T_x^{-1}) u ] \pa_u,
 \\ \nonumber
 K  = [ t^2 T_t^{-2} \Delta_t u + t x T_t^{-1} T_x^{-1} \Delta_x u + \frac{1}{4} 
x^2 T_x^{-2} u +
 \\ \nonumber
 + t ( T_t^{-2} - \frac{1}{2} T_t^{-1} T_x^{-1} ) u - \frac{1}{16} \sigma_x^2 
T_x^{-2} u ] \pa_u.
 \eea
 We see that the translations $P_0$ and $P_1$ involve only discrete derivatives, 
and the Galilei transformation $B$, dilation  $D$ and expansions $K$ involve 
explicit shifts to other points of the lattice. 
 
 Floreanini et. al. \cite{18,19} obtained an equivalent result for the heat 
equation (\ref{3.1}) where $\Delta_t \eq \Delta_t^-$ and $\Delta_{xx} \eq 
\Delta_{xx}^{--}$, i.e. they used "down derivatives", with for instance 
$\Delta_t^- = \frac{1-T_t^{-1}}{\sigma_t}$.
 
 In both cases shifts to a finite number of points are involved. Quite recently 
it was shown \cite{13c},\cite{25} that if symmetric discrete derivatives are 
used, e.g.
 \be \label{3.4}
 \Delta_x^s = \frac{T_x - T_x^{-1}}{2 \sigma_x}
 \ee
then the number of points involved in the symmetries, other than translations, 
will be infinite. Here we will stick with right derivatives, as in (\ref{3.2}) 
and (\ref{3.3}).

For Lie point symmetries as used in Ref. \cite{10,12} the definition of discrete 
derivative is immaterial. One simply considers relations between points in the 
lattice. To facilitate the comparison between "discrete" evolutionary and Lie 
point symmetries, we shall here consider the following heat equation and lattice
\bea \label{3.5}
\frac{u_{m,n+1}-u_{m,n}}{t_{m,n+1}-t_{m,n}} = \frac{u_{m+2,n} - 2 u_{m+1,n} + 
u_{m,n}}{(x_{m+1,n} - x_{m,n} )^2}
\\ \label{3.6}
x_{m+2,n} - 2 x_{m+1,n} + x_{m,n} = 0, \qquad x_{m,n+1} - x_{m,n} = 0
\\ \nonumber
t_{m+1,n} - t_{m,n} = 0, \quad t_{m,n+1} - t_{m,n} = c ( x_{m+1,n} - x_{m,n} 
)^2,
\eea
where $c$ is a constant. The symmetry algebra coincides with the one obtained 
earlier \cite{12} using a symmetric derivative on the right hand side of 
eq.(\ref{3.5}). It is spanned by 
\be \label{3.7}
\hat P_0 = \pa_t, \quad \hat P_1 = \pa_x, \quad \hat D = x\pa_x + 2 t \pa_t, 
\quad \hat W = u \pa_u, \quad \hat S = S(x,t) \pa_u
\ee
where $S(x,t)$ is a solution of the system (\ref{3.5}),(\ref{3.6}) and $\hat S$ 
represents the linear superposition principle.  The lattice equations 
(\ref{3.6}) can easily be solved and we have
\be \label{3.8}
x = \sigma_x m + x_0, \qquad t = c \sigma_x^2 n + t_0
\ee
where $\sigma_x$, $x_0$ and $t_0$ are integration constants that are not apriori 
fixed.

Let us now consider some nontrivial examples.

\subsection{ Reductions by Lie point symmetries}
\bigskip
{\bf 1. Translationally invariant solutions}
\bigskip

A solution of the system (\ref{3.5}), (\ref{3.6}), invariant under a translation 
generated by $\hat P_0 - a \hat P_1$ will have the form
\be \label{3.9}
u(x,t) = u(z), \qquad z = x + a t
\ee
The lattice equations (\ref{3.6}) and the heat equation (\ref{3.5}) reduce to
\bea \label{3.10}
z_{m+1,n} - 2 z_{m,n} + z_{m-1,n} = 0, \quad z_{m,n+1} - z_{m,n} = a c ( z_{m,n} 
- z_{m-1,n} )^2
\\ \label{3.11}
u(z_{m,n+1}) - u(z_{m,n}) = c [ u(z_{m+2,n}) - 2 u(z_{m+1,n}) + u( z_{m,n} ) ].
\eea
The solution of eq.(\ref{3.10}) is 
\be \label{3.12}
z_{m,n} = A ( m+ A a c n ) + z_0,
\ee
so $z_{m,n}$ really depends on just one label $N=m+Aacn$ ( $A$ and $z_0$ are 
integration constants).  If the reduced equation (\ref{3.11}) is supposed to be 
a difference equation on some lattice, the label $N$ must vary over integer 
values ($N$ simply enumerates different points $z_0$, $z_{\pm 1}$, $z_{\pm 2}$, 
..., independently of their spacing).  This implies that the constants $A$, 
$c$ and $a$ are constrained by the requirement 
\be \label{3.13}
A a c = k, \qquad k \in Z.
\ee

Eq.(\ref{3.11}) can be written as
\be \label{3.14}
u(z + a c A^2) - u(z) = c [ u(z + 2 A) - 2 u(z + A) + u(z) ].
\ee

The general solution of eq.(\ref{3.14}) can be written in the form
\be \label{3.15}
u(x,t) = c_1 e^{\alpha z} + c_2,
\ee
where the constant $\alpha$ is determined by the condition
\be \label{3.16}
e^{\alpha a c A^2} - 1 = c [ e^{2 \alpha A} - 2 e^{\alpha A} + 1 ].
\ee
Eq.(\ref{3.15}) also represents a translationally invariant solution of the 
continuous heat equation for $\alpha=a$. In the continuous limit we have $A = 
\sigma_x \ra 0$ and to order $\sigma_x^2$ eq.(\ref{3.16}) reduces to $\alpha = 
a$.

In general eq.(\ref{3.16}) is a trascendental equation for $\alpha$ and 
eq.(\ref{3.14}) determines the solution $u(z)$ at a point $z+acA^2$ on the $z$ 
line in terms of $u(z)$ at three given points.  As stated above, this will be a 
point on the same lattice if we choose $acA^2 = kA$ with $k$ integer.
If (\ref{3.13}) is satisfied, then eq.(\ref{3.16}) is an algebraic one for 
$v=e^{\alpha A}$:
\be \nonumber
v^k - 1 = c [ v^2 - 2 v +1 ].
\ee
In particular, for $k=1$ eq.(\ref{3.14}) is a three point difference equation 
and we have 
\be \label{3.17}
\alpha A = ln \frac{c + 1}{c}.
\ee
\bigskip 

\noindent {\bf 2. Reduction by dilation $\hat D$.}
\bigskip

A scaling invariant solution will have the form
\be \label{3.18}
u = u(z), \qquad z = x t^{-\frac{1}{2}}.
\ee
Eq.(\ref{3.6}) imply
\be \label{3.19}
z_{m+1,n} - 2 z_{m,n} + z_{m-1,n} = 0, \quad z_{m,n+1} = 
\frac{z_{m,n}}{\sqrt{1+c(z_{m,n}-z_{m-1,n})^2}}.
\ee
Solving eq.(\ref{3.19}), we have
\be \label{3.20}
z_{m,n} = \frac{m-m_0}{\sqrt{x(n-n_0)}},
\ee
where $m_0$ and $n_0$ are constants. E.(\ref{3.20}) for $z_{m,n}= const.$ 
determines a parabola in the ($m$,$n$)-plane, so $u_{m,n} \eq u(z_{m,n})$ is 
constant along each parabola. The discrete heat equation is reduced to the 
equation (\ref{3.11}).
Choosing some reference point $z=z_{m,n}$ we obtain an 
equation that can be written as
\be \label{3.22}
u(z \frac{\gamma_{n+1}}{\gamma_n}) - u(z) = c [ u(z + 2 \gamma_n) - 2 u(z + 
\gamma_n) + u(z) ], \quad \gamma_n = \frac{1}{\sqrt{c(n-n_0)}}
\ee
Even though we are not able to solve eq.(\ref{3.22}) analytically, we see that a 
reduction has occurred.  Eq.(\ref{3.22}) involves one independent variable $z$, 
rather than two.

For instance we can take $c=1$, $m_0=n_0=0$ in eq.(\ref{3.20}) which then 
reduces to
\be \label{3.23}
z_{m,n+1} = z_{m,n} \sqrt{\frac{n}{n+1}}.
\ee
For any fixed value of $m$ we need to give the values of $u(z)$ in a set of 
equally spaced points $z$, $z \pm \gamma_n$, $z \pm 2 \gamma_n$, ... .
Eq.(\ref{3.22})then determines $u(z)$ at irrationally spaced points (\ref{3.23}) 
along the same line. 

Thus we have reduced to an equation with one independent variable only, but it 
is not a difference equation, rather a {\sl difference - delay} one.

\subsection{Reduction by discrete evolutionary symmetries}
We rewrite eq.(\ref{3.1}) as
\bea \label{3.24}
u_{m,n+1} - u_{m,n} = c ( u_{m+2,n} - 2 u_{m+1,n} + u_{m,n}),\quad c=
\frac{\sigma_t}{\sigma_x^2}
\\ \label{3.25}
x_{m+1,n} - x_{m,n} = \sigma_x, \quad 
t_{m+1,n} - t_{m,n} =0,
\\ \nonumber
x_{m,n+1} - x_{m,n} = 0, \qquad t_{m,n+1} - t_{m,n} = \sigma_t
\eea
and consider reductions of this system by some of the "discrete" evolutionary 
symmetries.
\bigskip

\noindent{\bf 1. Translationally invariant solutions}
\bigskip

The commuting flow corresponding to a general translation is given by (see $P_0$ 
and $P_1$ in eq.(\ref{3.3})):
\be \label{3.26}
\frac{d u_{m,n}}{d\lambda} = \frac{u_{m,n+1} - u_{m,n}}{\sigma_t} - a 
\frac{u_{m+1,n} - u_{m,n}}{\sigma_x},
\ee
and an invariant solution will satisfy
\be \label{3.27}
u_{m,n+1} - u_{m,n} = a \frac{\sigma_t}{\sigma_x} (u_{m+1,n} - u_{m,n} ).
\ee
Together with eq.(\ref{3.24}) this implies
\be \label{3.28}
a (u_{m+1,n} - u_{m,n} ) = \frac{1}{\sigma_x} ( u_{m+2,n} - 2 u_{m+1,n} + 
u_{m,n} ).
\ee
This is a linear 
three point difference equation in $m$. Its general solution 
is
\be \label{3.29}
u_{m,n} = A(n) + B(n) (1 + a \sigma_x)^m.
\ee
Substituting back into the heat equation (\ref{3.24}) we find
\be \label{3.30}
A(n+1) = A(n), \qquad B(n+1) = (1 + a^2 \sigma_t ) B(n).
\ee
The general solution of the system (\ref{3.24}, \ref{3.27}) hence is
\be \label{3.31}
u_{m,n} = c_1 ( 1 + a^2 \sigma_t )^n ( 1 + a \sigma_x )^m + c_2.
\ee
Using the lattice conditions (\ref{3.25}) we can rewrite this as
\be \label{3.32}
u(x,t) = c_1 ( 1 + a \sigma_x)^{\frac{x}{\sigma_x}} (1 + a^2 
\sigma_t)^{\frac{t}{\sigma_t}} + c_2.
\ee
This is not the same solution (\ref{3.15}) obtained using translations as point 
symmetries. The two translationally invariant solutions only coincide in the 
continuous limit $\sigma_x \ra 0$, $\sigma_t \ra 0$.
\bigskip

\noindent {\bf 2. Reduction by dilations}
\bigskip

We shall use the operator $ D - (1 - \frac{1}{2} T_x^{-1})  W$ (see 
eq.(\ref{3.3})) to perform the reduction. Thus, we solve eq.(\ref{3.24}) 
together with the self-similarity condition
\be \label{3.33}
2 t_{m,n} \frac{u_{m,n} - u_{m,n-1}}{\sigma_t} + x_{m,n} \frac{u_{m,n} - 
u_{m-1,n}}{\sigma_x} = 0.
\ee
We solve eq.(\ref{3.33}) for $u_{m,n-1}$ and shift eq.(\ref{3.24}) down in $n$, 
i.e. replace $n$ by $n-1$ everywhere. Substituting for $u_{m+2,n-1}$, 
$u_{m+1,n-1}$ and $u_{m,n-1}$ we obtain the reduced equation
\bea \label{3.34}
2c(n+1)[u_{m+2,n}-2 u_{m+1,n} + u_{m,n}] + m [ 
c(u_{m+2,n}-u_{m+1,n})-\\ \nonumber -2c(u_{m+1,n}-u_{m,n}) 
+(c+1)(u_{m,n}-u_{m-1,n})] =0.
\eea
In eq.(\ref{3.34}) $m$ is a variable, $n$ is a parameter. The continuous limit 
is obtained by putting $x=\sigma_x m$, $t=\sigma_t n$, and taking $\sigma_x$ and 
$\sigma_t$ to zero with $c=\frac{\sigma_t}{\sigma_x^2}$ finite. We obtain
\be \label{3.35}
2 t u_{xx} + x u_x = 0.
\ee
This is indeed the condition for invariance under dilations generated by $\hat D 
$ of eq. (\ref {3.7}).
Eq.(\ref{3.35})is easily solved, eq.(\ref{3.34}) is more difficult to deal with.

We start by integrating eq.(\ref{3.34}) once. The equation involves four values 
of $m$. We reduce the number to three by putting
\be \label{3.36}
v_{m,n} \eq \frac{u_{m+1,n} - u_{m,n}}{\sigma_x}
\ee
and obtain
\be \label{3.37}
2c(n+1)(v_{m+1,n}-v_{m,n})+m[cv_{m+1,n}-2cv_{m,n}+(c+1)v_{m-1,n}]=0
\ee
This is a linear ordinary three point difference equation with variable 
coefficients.

In order to solve it we use a discrete Fourier transform (often also called the 
Z-tranform), i.e. introduce a generating function
\be \label{3.38}
G_n(z) = \sum_{m=-\infty}^{\infty} v_{m,n} z^m
\ee
with
\be \label{3.39}
v_{m,n} = \frac{1}{2 \pi i} \oint_{C_1} \frac{dz G_n(z)}{z^{m+1}} ,
\ee
where the contour $C_1$ is the unit circle in the complex $z$ plane. 
Eq.(\ref{3.37}) then implies that $G_n(z)$ satisfies
\be \label{3.40}
2c(n+1)(\frac{1}{z}-1)G_n 
+[c(G_{n,z}-\frac{G_n}{z})-2czG_{n,z}+(c+1)(z^2G_{n,z}+zG_n)]=0.
\ee
Eq.(\ref{3.40}) is easily integrated and we obtain
\be \label{3.41}
G_n(z) = \gamma_n \frac{(z-z_1)^n (z-z_2)^n}{z^{2n+1}},
\ee
where $\gamma_n$ is an integration constant and the complex numbers $z_{1,2}$ 
are
\be \label{3.42}
z_{1,2} = \frac{c \pm i \sqrt{c}}{c+1}.
\ee
Since we have $c>0$, $z_{1,2}$ lie inside the unit circle and eq.(\ref{3.39}) 
implies
\be \label{3.43}
v_{n,m} = \frac{1}{2 \pi i} \gamma_n \oint_{C_1} \frac{dz (z-z_1)^n (z - 
z_2)^n}{z^{m+2n+2}}.
\ee
The dependence on $m$ in eq.(\ref{3.43}) is explicit, the dependence of 
$\gamma_n$ on $n$ must still be determined. To do this we introduce the notation
\be \label{3.44}
v_{m,n} = \gamma_n I_{N,n}, \qquad N = m + 2 n + 2
\ee
and sustitute into eq.(\ref{3.33}) (the condition for dilational invariance). We 
also use $x_{m,n} = \sigma_x m$, $t_{m,n} = \sigma_t n$ and obtain
\be \label{3.45}
2n(u_{m,n}-u_{m,n-1})+m(u_{m,n}-u_{m-1,n})=0.
\ee
In order to introduce $v_{m,n}$ into eq.(\ref{3.45}) we first take the variation 
(discrete derivative) with respect to $m$ and then obtain
\be \label{3.46}
2n(v_{m,n}-v_{m,n-1}) + (m+1)v_{m,n}-mv_{m-1,n}=0.
\ee
Substituting the expression (\ref{3.44}) for $v_{m,n}$ we obtain an equation for 
$\gamma_n$, namely
\be \label{3.47}
\gamma_n[(N-1) I_{N,n}-(N-2n-2) I_{N-1,n}] -2n\gamma_{n-1} I_{N-2,n-1}=0.
\ee
Eq.(\ref{3.47})must hold for all values of $N$ and $\gamma_n$ is independent of 
$N$. The quantity $I_{N,n}$ is defined by the integral in eq.(\ref{3.43}) and 
can be evaluated using the residue theorem. For general $N$ and $m$ this is not 
very illuminating. For low values of $N$ and $n \ge0$ the results are quite 
simple, for instance
\bea \nonumber
I_{1,n} = & (z_1 z_2)^n = (\frac{c}{c+1})^n,
\\ \nonumber
I_{0.n} = & 0,
\\ \nonumber
I_{-1,n-1} = & 0,
\\ \nonumber
I_{2,n} = & - 2n (\frac{c}{c+1})^n,
\\ \nonumber
I_{3,n} = & n (\frac{c}{c+1})^{n-1} \frac{(2n-1)c-1}{c+1},
\\ \nonumber
I_{4,n} = & -\frac{2}{3}n(n-1) (\frac{c}{c+1})^{n-1} \frac{(2n-1)c+3}{c+1}.
\eea
Substituting these values into eq.(\ref{3.47}) we obtain
\be \label{3.48}
\gamma_n = (c+1)^n \gamma_0,
\ee
which can be shown to be valid for all integer values of $n$, both positive and 
negative.  Finally, the $m$ variation $v_{m,n}$ of the dilationally invariant 
solution of the discrete heat equation (see eq.(\ref{3.36})) satisfies 
\be \label{3.49}
v_{m,n} = \gamma_0 (c+1)^n \frac{1}{2 \pi i} \oint_{C_1} \frac{dz (z-z_1)^n 
(z-z_2)^n}{z^N},
\ee
where $C_1$ is the unit circle and $z_{1,2}$ are defined in eq.(\ref{3.42}).

The continuous limit of this self similar solution is
\be \label{3.50}
v(x,t) = \frac{d}{dx} u(x,t) = \frac{\gamma_0}{\sqrt{t}} e^{\frac{-x^2}{4t}}
\ee
which satisfies the continuous limit of eq.(\ref{3.33}), namely $2tu_t + x u_x = 
0$, which implies
\be \label{3.51}
2tv_x+xv=0.
\ee

\section{Symmetry Reduction for the Discrete Toda lattice}

A recent article was devoted to a hierarchy of nonlinear integrable difference 
equations associated with a discrete Schr\"odinger spectral problem \cite{23}. 
The hierarchy includes the discrete Toda and discrete Volterra lattice 
equations. All equations in the hierarchy involve two independent variables: 
discrete space and discrete time.

In this section we shall use Lie point symmetries and "discrete"  evolutionary 
symmetries to perform symmetry reduction for the simplest equation in the 
hierarchy,  namely the Discrete Time Toda Lattice (DTTL) itself: 
\bea \label{4.1}
e^{u_{n,m} - u_{n,m+1}} - e^{u_{n,m+1} - u_{n,m+2}} 
= \alpha^{2} ( 
e^{u_{n-1,m+2} - u_{n,m+1}} - e^{u_{n,m+1} - u_{n+1,m}} ),
\eea
( $\alpha$ is a constant).

\subsection{Reduction by Lie point symmetries}

We complement the DTTL (\ref{4.1}) by the lattice equations:
\bea \label{4.2}
x_{n+1,m} - 2 x_{n,m}  + x_{n-1,m} & =  0,
\qquad x_{n,m} & =  x_{n,m+1},
\\ \nonumber
t_{n,m+1} - 2 t_{n,m} + t_{n,m-1}& = 0, \qquad
t_{n,m} & =  t_{n+1,m}.
\eea
The solution of eq.(\ref{4.2}) is 
\be \label{4.3}
x_{n,m} = \sigma_x n, \qquad t_{n,m} = \sigma_t m,
\ee
where the integration constants $\sigma_x$ and $\sigma_t$ represent the lattice 
spacings and we have set the two further "initial value" constants $x_{0,m}$ and 
$t_{n,0}$ equal to zero.

The Lie algebra of the Lie point symmetry group of the difference system 
(\ref{4.1}) and (\ref{4.2}) is spanned by 
\be \label{4.3a}
P_0 = \partial_t, \quad P_1 = \partial_x, \quad D_0 = t \partial_t, \quad D_1 = 
x \partial_x, \quad W = \partial_u.
\ee

Let us look separately at reductions by translations and by dilations.
\bigskip

\noindent{\bf A. Translationally invariant solutions.}
\bigskip

A general translationally invariant solution has the form 
\bea \label{4.4}
u_{n,m} \eq u(x_{n,m},t_{n,m}) = u(\xi_{n,m}), \\ \nonumber
 \xi_{n,m} = x_{n,m} + a t_{n,m} = \sigma_x n + a \sigma_t m.
\eea

Eq.(\ref{4.1}) reduces to an equation for $u(\xi)$ evaluated at five points 
$\xi$, namely
\bea \label{4.5}
\xi_{n,m} \eq & \xi, \quad \xi_{n,m+1}  & =  \xi + a \sigma_t, \quad  
\xi_{n,m+2} = 
\xi + 2 a \sigma_t,
\\ \nonumber  
\xi_{n+1,m} = & \xi + \sigma_x, \quad
 \xi_{n-1,m+2} & = 
  \xi - \sigma_x + 2 a \sigma_t.
\eea
As in the case of the heat equation (see eq. (\ref{3.13},\ref{3.14})) the 
reduced equation will be an ordinary difference equation on a lattice only if 
$\xi_{n,m}$ is a function of one discrete label. This occurs if we put
\be \label{4.6}
a \sigma_t = k \sigma_x, \qquad k \in Z
\ee
in eq.(\ref{4.4}). In particular, if we choose $k=1$ in eq.(\ref{4.6}) we obtain 
a three point difference equation (since $\xi_{n,m+1} = \xi_{n+1,m} = 
\xi_{n-1,m+2}$). The right hand side of eq.(\ref{4.1}) vanishes identically and 
the left hand side implies 
\be \label{4.7}
u_{n,m} - 2 u_{n,m+1} + u_{n,m+2} = 0.
\ee
The general solution of eq.(\ref{4.7}) is
\be \label{4.8}
u_{n,m} = f(n)  m + g(n).
\ee
Substituting back into eq.(\ref{4.1}) we obtain the general translationally 
invariant solution of eq.(\ref{4.1}) as 
\be \label{4.9}
u_{n,m} = A  n ( n + m ) + B m + C n + D,
\ee
where $A$, $B$, $C$ and $D$ are constants.
\bigskip

\noindent {\bf B. Reduction by dilations}
\bigskip

A solution invariant under dilations generated by $D_0 - \beta D_1$ will have 
the form
\bea \label{4.10}
u(x_{n,m},t_{n,m}) = u(\xi_{n,m}) \quad 
\xi_{n,m} = x_{n,m} t_{n,m}^{\beta} = \sigma_x \sigma_t^{\beta} n m^{\beta}
\equiv\xi.
\eea
Substituting into the DTTL (\ref{4.1}) we obtain a nonlinear equation involving 
$u(\xi)$ evaluated at $5$ points:
\be \label{4.11}
e^{u(\xi) - u(\xi_1)} - e^{u(\xi_1) - u(\xi_2)} = \alpha^{2} ( 
e^{u(\xi_3) - u(\xi_1)} - e^{u(\xi_1) - u(\xi_4)} ),
\ee
with $\xi$ as in (\ref{4.10}) and
\bea \label{4.12}
\xi_1 = \xi (\frac{m+1}{m})^{\beta}, \quad \xi_2 & =& \xi 
(\frac{m+2}{m})^{\beta},
\\ \nonumber  \xi_3 = \xi (\frac{m+2}{m})^{\beta} - \sigma_x \sigma_t^{\beta} 
(m+2)^{\beta},
\xi_4 & =& \xi + \sigma_x \sigma_t^{\beta} m^{\beta}
\eea
Eq. (\ref {4.11}) is a {\sl dilation - delay} equation.

\subsection{Discrete Evolutionary Symmetries for the DTTL}

The DTTL (\ref{4.1}) can be written as a system of two equations \cite{23} as
\bea \label{4.13}
a_{n,m+1} - a_{n,m} = & \alpha ( b_{n,m+1} - b_{n+1,m} ) 
\frac{\pi_{n,m+1}}{\pi_{n+1,m}},
\\ \label{a12}
b_{n,m+1} - b_{n,m} = & \alpha (  \frac{\pi_{n-1,m+1}}{\pi_{n,m}} - 
\frac{\pi_{n,m+1}}{\pi_{n+1,m}}).
\eea
where we have
\bea \label{4.14}
\pi_{n,m} =& a_{n,m} \pi_{n+1,m}, \qquad \pi_{n,m} = e^{u_{n,m}},
\\ \nonumber
\pi_{n,m} =& \Pi_{j=n}^{\infty} a_{j,m}.
\eea
As in the case of the Toda lattice itself, both isospectral and nonisospectral 
"discrete" evolutionary symmetries exist \cite{9,13,23} and we shall consider 
both 
of them.

The simplest nontrivial isospectral (generalized) symmetry is given by
\bea \label{4.15}
(a_{n,m})_\epsilon & =& a_{n,m} [ a_{n-1,m} - a_{n+1,m} + b_{n,m}^2 - 
b_{n+1,m}^2 
]
\\ \nonumber
(b_{n,m})_\epsilon & =& a_{n-1,m} [ b_{n,m} + b_{n-1,m} ] - a_{n,m} [ b_{n+1,m} 
+ 
b_{n,m} ].
\eea

We assume $a_{n,m} \ne 0$.  Symmetry reduction amounts to solving 
eq.(\ref{4.15}) for $(a_{n,m})_\epsilon = (b_{n,m})_\epsilon = 0$. 
Eq.(\ref{4.15}) can be once integrated to yield
\bea \label{4.16}
a_{n-1,m} + a_{n,m} + b_{n,m}^2 = A_m
\\ \nonumber
a_{n,m} ( b_{n+1,m} + b_{n,m} ) = B_m.
\eea
We can eliminate $b_{n,m}$ from (\ref{4.16}) to obtain 
\be \label{4.17}
a_{n,m} ( {\sqrt{A_m - a_{n,m} - a_{n+1,m}}} + {\sqrt{A_m - a_{n-1,m} - 
a_{n,m}}} ) = B_m.
\ee
Eq.(\ref{4.17}) can be viewed as a discrete analog of the equation for elliptic 
functions. It is an integrable difference equation in one variable $n$ with $m$ 
as a fixed paramether. The functions $A_m$ and $B_m$ must be determined by 
putting 
the solution of eq.(\ref{4.16}) into (\ref{4.13}).

The simplest nonisospectral symmetry is given by
\bea \label{4.18a}
(a_{n,m})_\epsilon = & a_{n,m} [ ( 2 n + 2 m + 3) b_{n+1,m} - ( 2 n + 2 m - 1) 
b_{n,m} ] 
\\ \label{4.18b}
(b_{n,m})_\epsilon = & b_{n,m}^2 - 4 + 2 [ ( n + m + 1) a_{n,m} - ( n + m -1 ) 
a_{n-1,m} ],
\eea
( see eq. (64) in Ref.\cite{23} with $k=0$, $\alpha = 1$ ).

Again, we must solve the equations $(a_{n,m})_\epsilon = (b_{n,m})_\epsilon = 
0$. We eliminate $b_{n,m}$ from these two equations and obtain
\bea \label{4.19}
(2n + 2m +3)^2 (n+m+2) a_{n+1,m}  + 
\\ \nonumber
- [ (n+m)(2n+2m+3)^2 + (n+m+1)(2n+2m-1)^2 ] a_{n,m} + 
\\ \nonumber
+ (n+m-1)(2n+2m-1)^2 a_{n-1,m} = 16 ( 2n+2m+1).
\eea

One solution of the corresponding homogeneous linear ordinary difference 
equation (in which $n$ is the independent variable, $m$ a parameter) can be 
guessed, namely
\be \label{4.20}
a_{n,m}^1 = \frac{1}{(n+m)(n+m+1)}.
\ee
The general solution of the inhomogeneous equation is constructed in the form
\be \label{4.21}
a_{n,m} = a_{n,m}^1 \beta_{n,m}.
\ee
We define
\be \label{4.22}
c_{n,m} = \beta_{n+1,m} - \beta_{n,m}.
\ee 
Substituting (\ref{4.21}) into (\ref{4.19}) we obtain a first order equation for 
$c_{n,m}$, namely
\be \label{4.23}
c_{n+1,m} = \frac{(n+m+2)(2n+2m+1)^2}{(n+m+1)(2n+2m+5)^2} c_{n,m} + 16 
\frac{(n+m+2)(2n+2m+3)}{(2n+2m+5)^2}.
\ee
A particular solution of the homogeneous part of eq.(\ref{4.23}) is 
\be \nonumber
c_{n,m} = 2 (n + m+ 1)
\ee
which implies $\beta_{n,m} = n ( n+m+1)$. Finally, the general solution of 
eq.(\ref{4.19}) is
\be \label{4.24}
a_{n,m} = \frac{1}{(n+m)(n+m+1)} \{ A(m) + \frac{B(m)}{(2n+2m+1)^2} + n(n+2m+1) 
\}
\ee
where $A(m)$ and $B(m)$ are integration constants.
Returning to (\ref{4.18b}) for $(b_{n,m})_\epsilon = 0$ we find
\be \label{4.25}
b_{n,m} = \frac{4 B(m)}{(2n+2m-1)(2n+2m+1)}.
\ee
Together, eq.(\ref{4.24}) and (\ref{4.25}) represent the general solution of the 
invariance condition $(a_{n,m})_\epsilon = (b_{n,m})_\epsilon = 0$ in 
eq.(\ref{4.18a}, \ref{4.18b}).

The functions $A(m)$ and $B(m)$ must be determined by substituting $a_{n,m}$ and 
$b_{n,m}$ into the DTTL of eq. (\ref{4.13}) (for $\alpha=1$).  This can be done; 
 the result is:
\be \label{4.26}
B(m) = 0, \qquad A(m) = m ( m + 1 )
\ee
and finally
\be \label{4.27}
a_{n,m} = 1, \qquad b_{n,m} = 0.
\ee
The result is somewhat disappointing since the invariant solution is one in 
which the fields are equal to their asymptotic values for all $n$ and $m$.

In order to obtain more interesting solutions, higher symmetries must be 
considered. Those will however be nonlocal and that is beyond the scope of
 the present article.

\section{Conclusions}

We have shown how to use symmetries of difference equations in two 
variables, either Lie point or "discrete" evolutionary ones, to construct 
solutions by carrying out a symmetry reduction.

We are confronted with two different situations. In the case of Lie point 
symmetries the symmetry variables, as opposed to the case of partial 
differential equations, do not always reduce the equation to a difference 
equation in a space of lower dimension. We have to impose a further constraint 
to be able to do so.  Moreover in some cases, among them that of dilations, the 
system reduces to a dilation - delay equation which is difficult to solve.

The situation is different for "discrete" evolutionary symmetries.  These 
symmetries exist only when the discrete system is integrable and the symmetries 
do not act on the lattice.  The symmetry reduction can always be carried out, 
but the obtained reduced equation can be nonlinear and difficult to solve (this 
is also true in the continuous case).

\subsection*{Acknowledgements}
We are grateful for the support of the Isaac Newton Institute for Mathematical 
Sciences during the Integrable Systems Programme where part of this 
research was carried out. The research reported in this article was partly 
supported by research grants from NSERC of Canada, FCAR du Quebec, NATO and the 
Exchange agreement between Universit\'e de Montr\'eal and Universit\'a Roma Tre.

 
 \end{document}